\documentclass[journal]{IEEEtran}
\ifCLASSINFOpdf
  \usepackage[pdftex]{graphicx}
   \graphicspath{{../figures/}{../jpeg/}}
\else
\fi

\usepackage[cmex10]{amsmath}
\usepackage{amssymb}
\usepackage{bm}

\usepackage{algpseudocode}
\usepackage{algorithm}
\usepackage{array}
\usepackage{url}
\usepackage[noadjust]{cite}

\usepackage{fancyhdr}
\usepackage{amsfonts}
\usepackage{xfrac}
\usepackage{color}
\usepackage{pgfplots}
\pgfplotsset{compat=newest}           
\pgfplotsset{compat=1.16}
\usepackage{enumerate}

\usepackage{cite}
\usepackage{mathrsfs}
\usepackage{subcaption}
\usepackage{stfloats}

\usepackage{makecell}
\usepackage{comment}

\usepackage{afterpage}
\usepackage{tikz}
\usepackage{mathdots}
\usepackage{yhmath}
\usepackage{cancel}
\usepackage{color}
\usepackage{siunitx}
\usepackage{array}
\usepackage{multirow}
\usepackage{textcomp}
\usepackage{gensymb}
\usepackage{tabularx}
\usepackage{booktabs}
\usepackage{graphicx}
\usepackage{xcolor}

\allowdisplaybreaks

\usetikzlibrary{decorations.markings,decorations.pathmorphing}

\begin{document}
\title{Empowering Programmable Wireless Environments with Optical Anchor-based Positioning}

    
    
    

\author{Dimitrios Tyrovolas,~\IEEEmembership{Member,~IEEE}, Dimitrios Bozanis,~\IEEEmembership{Student Member,~IEEE}, \\Sotiris A. Tegos,~\IEEEmembership{Senior Member,~IEEE}, Vasilis K. Papanikolaou,~\IEEEmembership{Member,~IEEE},\\ Panagiotis D. Diamantoulakis,~\IEEEmembership{Senior Member,~IEEE},  Christos K. Liaskos,~\IEEEmembership{Member,~IEEE}, \\ Robert Schober,~\IEEEmembership{Fellow,~IEEE}, and George K. Karagiannidis,~\IEEEmembership{Fellow,~IEEE}

\thanks{D. Tyrovolas, D. Bozanis, S. A. Tegos and P. D. Diamantoulakis are with the Aristotle University of Thessaloniki, 54124 Thessaloniki, Greece (tyrovolas@auth.gr, dimimpoz@ece.auth.gr, tegosoti@auth.gr, padiaman@auth.gr).}
\thanks{V. K. Papanikolaou and R. Schober are with the Friedrich-Alexander-University Erlangen-Nuremberg, 91058 Erlangen, Germany (vasilis.papanikolaou@fau.de, robert.schober@fau.de).}
\thanks{C. K. Liaskos is with University of Ioannina, Greece and with Foundation for Research and Technology Hellas (FORTH), Greece (cliaskos@uoi.gr).}
\thanks{G. K. Karagiannidis is with the Aristotle University of Thessaloniki, 54124 Thessaloniki, Greece and with the Cyber Security Systems and Applied AI Research Center, Lebanese American University, Lebanon (geokarag@auth.gr).}

}
 
\maketitle
\begin{abstract}
The evolution toward sixth-generation (6G) wireless networks has introduced programmable wireless environments (PWEs) and reconfigurable intelligent surfaces (RISs) as transformative elements for achieving near-deterministic wireless communications. However, the enhanced capabilities of RISs within PWEs, especially as we move toward more complex electromagnetic functions by increasing the number of reflecting elements, underscore the need for high-precision user localization, since inaccurate localization could lead to erroneous configuration of RISs, which would then compromise the effectiveness of PWEs. In this direction, this paper investigates the integration of RISs and optical anchors within PWEs, emphasizing the crucial role of ultra-precise localization in unlocking advanced electromagnetic functionalities. Specifically, we present an in-depth analysis of various localization techniques, both RIS-based and RIS-independent, while introducing the concept of empowering PWEs with optical anchors for enhanced localization precision. Our findings highlight that accurate localization is essential to fully exploit the capabilities of RISs, paving the way for future applications. Through this exploration, we contribute to the advancement of PWEs in line with the ambitious goals of the 6G standards and improve the quality of service in next-generation wireless networks. 
\end{abstract}

\begin{IEEEkeywords}
Reconfigurable Intelligent Surfaces (RISs), Programmable Wireless Environments (PWEs), Light-emitting Reconfigurable Intelligent Surfaces (LERISs), Visible Light Positioning (VLP), Localization
\end{IEEEkeywords}

\section{Introduction}
The transition to sixth-generation (6G) wireless networks marks a significant advancement in communications, introducing a multi-band framework designed to harness each frequency band's unique benefits, thereby offering intelligence, unlimited connectivity, and a seamless fusion of the physical and digital realms. At the heart of this transition are the emerging 6G standards focused on fostering novel network capabilities including connected sensing, sustainable communications, and the integration of artificial intelligence (AI) into wireless networks. Aligned with these ambitions, the concept of \emph{programmable wireless environments (PWEs)} was introduced as a transformative approach for converting propagation environments into intelligent spaces that can deliver wireless services with near-deterministic accuracy \cite{liaskos2018}. Achieving this involves the utilization of reconfigurable intelligent surfaces (RISs), programmable metasurfaces consisting of individually controllable reflecting elements, capable of manipulating electromagnetic waves for varied functions such as beam steering, splitting, and absorption \cite{basar2019}. Macroscopically, RISs are thin, and planar devices resembling tiles that interact with impinging electromagnetic waves in a software-configurable manner, while microscopically, they serve as hardware platforms enabling the software-defined transformation of its elements' reflection coefficient. This transformation process involves configuring the RIS by precisely tuning embedded active components such as PIN diodes, resulting in the formation of a desired scattering profile that corresponds to precise EM functionalities. This control over wireless propagation translates into significant improvements in quality of service (QoS) and network capabilities, enhancing the efficiency of electromagnetic spectrum use and reducing both time and cost associated with network upgrades, thus underscoring the critical role of RISs in 6G wireless network standardization efforts \cite{aryan2024}.

In the PWE paradigm, optimizing the RIS configuration of RISs is pivotal for enhancing network adaptability and performance, necessitating novel methodologies like AI and predictive analytics to ensure alignment with user needs and network demands \cite{AICui2021}. Among these, codebooks emerge as a notably practical method, comprising pre-calculated optimal RIS configurations for specific scenarios, thereby facilitating rapid adaptation to network changes without the complexity associated with iterative optimization or extensive data training \cite{direnzo2022codebook}. However, the enhanced capabilities of RISs within PWEs, especially as we pave the way towards more detailed electromagnetic functionalities through an increased number of reflecting elements, underscore the necessity for high-precision user localization. This need arises as any misalignment from inaccurate localization may result in the selection of incorrect functionalities from the codebook, thereby diminishing PWE effectiveness. Specifically, applications demanding precise wavefront shaping, such as extended reality-radio frequency (XR-RF) where RISs direct precise RF wavefronts to the receiver that are then translated to specific graphical outputs from the user's device, highlight the critical role of accuracy where small location estimation errors can significantly impair QoS \cite{XRRF}. Thus, the integration of methodologies for accurate localization is crucial for enabling the full utilization of RIS capabilities and customization to user-specific demands.

In this paper, we study the integration of RISs and light-emitting diodes (LEDs) as optical anchors within PWEs, highlighting their critical role in facilitating precise localization, which is crucial for unlocking advanced RIS functionalities expected in 6G networks. Specifically, we provide a thorough analysis of various localization methods, including both RIS-based and independent approaches, to reveal their capabilities and limitations within PWEs. This analysis highlights the essential synergy between RISs and optical technologies, and leads us to introduce the \emph{light-emitting reconfigurable intelligent surface (LERIS)}, i.e., an RIS equipped with four IR LEDs that are directly coordinated by the RIS controller, allowing the user equipment (UE) to accurately determine its position by utilizing the unique signal profiles emitted by the LERIS. Importantly, a LERIS uses optical technology for localization while continuing to manipulate RF wavefronts, thus maintaining its core functionality within the RF domain and significantly increasing the localization accuracy required for sophisticated electromagnetic operations. Finally, by highlighting the accuracy of optical anchor-based localization in enhancing the effectiveness of PWE through simulations, our work demonstrates that high-precision localization is critical for utilizing the advanced capabilities of RISs equipped with a large number of reflecting elements.

\section{A Roadmap to Precise Localization for PWEs}\label{sec:II}

\begin{table*}[t]
\caption{Comparison of Localization Methods}
\begin{center}
\setlength\tabcolsep{1pt}
\small
\begin{tabular}{|l|c|c|c|c|c|}
\hline
\textbf{Localization Method} & \textbf{\makecell{Robustness to \\ Multipath}} & \textbf{Low Latency} & \textbf{\makecell{Increased \\ Complexity}} & \textbf{Scalability} & \textbf{Data Privacy} \\
\hline
Beam-Scanning \cite{10100675} & & & \checkmark & & \checkmark \\
\hline
Diffusion \cite{9782100}& & \checkmark & & \checkmark & \checkmark \\
\hline
Gridding \cite{9456027}& & \checkmark & \checkmark & & \checkmark \\
\hline
RF-Anchors \cite{7438736}& & \checkmark & & \checkmark & \checkmark \\
\hline
Camera-Based \cite{10041749} & \checkmark & & \checkmark & & \\
\hline
Optical-Anchors \cite{10180234} & \checkmark & \checkmark & & \checkmark & \checkmark \\
\hline
\end{tabular}
\label{tab1}
\end{center}
\end{table*}

To ensure effective operation in all potential user scenarios within PWEs, the development of precise localization techniques becomes vital, especially as we advance the capabilities of RIS to facilitate intricate electromagnetic functions. To address this need, the development of locate-and-then-configure (LATC) schemes should incorporate advanced localization strategies to pinpoint and tailor RIS configurations from a comprehensive codebook, a data structure that maps RIS functionalities to specific RIS configurations. This intricate relationship between RIS functionalities and localization precision underscores the categorization of LATC schemes into those that are based on specific RIS functionalities and those based on external systems that perform localization, demonstrating the range of available methods for accurate localization within PWEs that also align with the dynamic requirements of 6G networks. The strengths and limitations of these methods in terms of robustness to multipath effects, latency, complexity, scalability, and data privacy are further detailed in Table \ref{tab1}.

\subsection{RIS-Based Localization}

Within the PWE framework, the strategic deployment of RISs throughout the environment unlocks new paths for enhancing localization through RF wavefront manipulation. Guided by codebooks, the operation of RISs lays the groundwork for creating specific localization functionalities that can outperform traditional systems, eliminating the necessity for direct line-of-sight (LoS) between access points (APs) and UE. Such flexibility allows the UE to employ angle of arrival (AoA), received signal strength (RSS), and time difference of arrival (TDoA) based methods for location determination exploiting RISs, with subsequent steps involving reporting their location to the network \cite{RISKarag2023}. Therefore, the unique capabilities of RISs necessitate the exploration of different functionalities designed for UE localization.

\begin{itemize}
    \item \textbf{Beam Scanning}: This functionality employs RISs to dynamically generate directional beams across the entire service area, aiming to detect UEs by systematically directing concentrated beams, thus bypassing the need for direct LoS between the AP and the UE \cite{10100675}. However, beam scanning involves a trade-off between the latency induced by the highly directional beams required for scanning the whole space, and faster but less precise location estimation provided by broader beams. Furthermore, as the PWE expands, beam scanning requires additional RISs with enhanced capabilities for coordinated beam steering over larger areas, compounded by the need for advanced control systems for beam management and synchronization to adapt in real-time to UE movement and environmental changes. This increases complexity, as ensuring coherent RIS operation requires strict calibration and alignment to maintain accurate coverage and avoid overlaps or gaps. Therefore, while beam scanning is an innovative functionality for UE localization in PWEs, its practicality is closely linked to managing the trade-off between scanning coverage and operational latency, underscoring the challenges of achieving optimal localization performance.    
 
    \item \textbf{Diffusion}: Within a PWE where multiple RISs are strategically deployed, we can facilitate accurate positioning by homogeneously diffusing electromagnetic waves with the UE to enable trilateration \cite{9782100}. The simplicity of this method allows the addition of more diffusing RIS units to seamlessly extend coverage and improve scalability without adding complexity to the system architecture or requiring costly hardware upgrades. However, this strategy faces challenges, such as the necessity for multiple RISs with a LoS to the UE, which increases deployment costs and operational complexity. Moreover, the effectiveness of diffusion depends on balanced signal distribution across the PWE, a process complicated by environmental variables and the physical placement of RISs, and is susceptible to multipath effects, indicating that the environmental conditions of the PWE have a critical impact on localization accuracy. Thus, while the diffusion approach offers a novel localization strategy within PWEs, its successful implementation depends on careful RIS placement and a comprehensive cost-benefit analysis.
    
    \item \textbf{Gridding}: The gridding functionality within the PWE framework leverages RISs to segment the service area into distinct grids, each characterized by unique RSS values, thereby directly correlating the UE position with the RSS value it encounters \cite{9456027}. However, this innovative approach requires high precision for optimal signal distribution across grids, a task that can prove challenging with a single RIS, and is further complicated by the necessity for significant RSS differentiation between adjacent grids for accurate position identification. Moreover, ensuring uniform gridding performance in dynamic environments requires extensive coordination among multiple RIS units and regular codebook updates, in response to environmental changes. Consequently, while gridding is considered a novel strategy for localization within PWEs, its successful deployment depends on overcoming the mentioned technical challenges.

\end{itemize}

\subsection{RIS-Independent Localization}

Despite the promising capabilities of RIS-based localization strategies, their full potential in PWEs is constrained by hardware limitations that prevent precise wavefront manipulation and multipath effects arising from uncontrolled transmissions that do not impinge upon the RISs, leading to reduced localization accuracy. Consequently, independent localization systems that are dedicated to accurate user positioning can prove critical for PWEs to dynamically and accurately adapt to user-specific needs, by enhancing the accuracy and robustness of RIS configurations in complex environments. To this end, it becomes imperative to explore independent systems to achieve effective localization, and thus PWE adaptability to UE dynamic requirements.

Building on this, RF anchors constitute a traditional independent localization system, leveraging four RF nodes to maintain direct LoS with the UE and employing RSS-based trilateration for localization. The inherent simplicity and direct communication of this approach facilitate fast and scalable localization, which is critical for real-time applications \cite{7438736}. However, its dependency on uninterrupted LoS paths poses significant challenges in environments with dynamic obstacles or dense structures. Additionally, while RF anchors operate independently of RISs, their presence in the environment can affect wave propagation and scattering, requiring carefully designed configurations to mitigate multipath effects and maintain localization accuracy, thereby illustrating the complexity of effectively integrating RF anchors into PWEs.

To fully unlock the potential of PWEs and align with the multi-band vision of 6G networks, pinpointing localization solutions that resist multipath effects and do not rely on specific RIS configurations is crucial. In this direction, optical wireless stands out as a viable technology, promising precise localization without RF spectrum interference issues or spectrum saturation. Specifically, advances in cameras and image processing techniques over the past decade have enabled highly accurate and robust localization capabilities \cite{10041749}. However, the practical application of camera-based methods faces several challenges, including the need for adequate illumination to ensure effectiveness, increased latency due to the complexity of processing image data, and privacy concerns related to the potential collection of sensitive information, as opposed to methods that process electromagnetic signals without capturing identifiable visual data. Moreover, camera-based localization exhibits significant hardware demands, like high-resolution cameras for various lighting conditions, powerful processors for real-time analysis, and adequate storage. Therefore, addressing these challenges necessitates a thorough assessment of camera-based localization requirements and the potential to create a system leveraging optical frequencies while mitigating the limitations of camera-based approaches.

\begin{figure*}[!t]
\centering{}\includegraphics[width=0.7\textwidth, trim=0cm 0cm 0cm 0.5mm, clip]{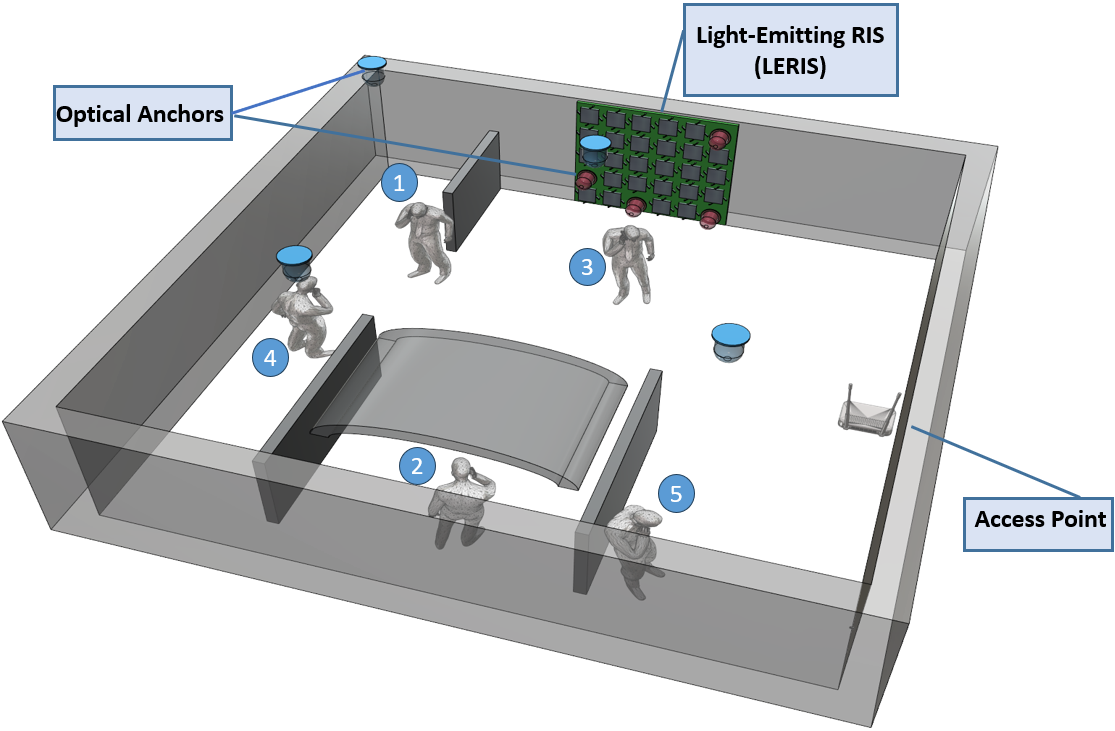}
\caption{Overview of a PWE empowered with optical anchors }\label{PWE_figure}
\end{figure*}

\section{Optical Anchors in PWEs}\label{sec:III}

In the effort to incorporate accurate localization within PWEs, the use of LEDs as optical anchors emerges as a strategic approach for configuring the available RISs that introduces several critical advantages. Unlike RF-based approaches, optical anchors leverage their primarily LoS channel to provide accurate location with minimal computation and robustness to multipath issues, thereby increasing the reliability of communication within PWEs \cite{10180234}. Furthermore, the use of ubiquitous lighting infrastructure meets the diverse and growing needs of PWEs, providing a scalable solution that is energy efficient and promotes sustainable network operation. In addition, the inherent privacy provided by optical anchor-based localization, which relies on signal strength and photodetector arrays rather than visual data collection, addresses privacy concerns that are increasingly relevant in the era of ubiquitous computing. Thus, formulating a well-defined PWE architecture that incorporates optical anchors can pave the way for a holistic localization strategy that meets the stringent requirements of 6G networks and ensures optimal performance through a persistent LoS between the UE and optical anchors.

\subsection{Integrating Optical Anchors within PWEs} 
Given that a LoS between optical anchors and the UE is essential for accurate localization, achieving comprehensive LED coverage within PWEs is a significant challenge. To address this, given the need to arrange the optical anchors to accommodate different lighting conditions and receiver orientations, we propose to deploy multiple LEDs on the ceiling and on the available RISs, as shown in Fig. \ref{PWE_figure}. These LEDs are coordinated by a central node using either contention-free multiple access protocols, wavelength division multiplexing with appropriate optical filters at the receivers to avoid inter-channel interference, or in a multiple-input multiple-output (MIMO) configuration. As a result, the network adapts to UE movement by combining the received information from multiple available optical anchors, improving localization accuracy through dynamic LED selection, and seamlessly adapting to the dynamic requirements of the PWE.

By leveraging the strategic placement of RISs to provide ubiquitous functionality, an innovative solution is the development of LERIS, which guarantees robust LoS links for all potential UE locations within the PWE. As shown in Figure \ref{PWE_figure}, this design involves equipping each RIS with four IR LEDs that are directly coordinated by the RIS controller. As each LED emits an identifiable signal to the receiver, it becomes possible for the UE to accurately determine its position and orientation with respect to the RIS by using the unique emitted optical signal profiles. This eliminates the dependency on ceiling optical anchors, which may be blocked. As a result, LERIS represents a seamless integration of optical and metasurface technologies, marking a significant advancement in the development of compact, efficient and accurate localization systems, perfectly aligned with the vision of PWEs.

Among the pivotal localization techniques facilitated by optical anchors is the RSS method, which necessitates the simultaneous illumination of the UE by at least four LEDs, laying the groundwork for trilateration. This approach entails minimal computational demand on the UE, highlighting its efficiency and simplicity as a localization strategy. Furthermore, by carefully placing the LEDs within the PWE, the UE orientation can also be determined, thus reducing the need for additional receiver-side devices to acquire orientation. However, particularly in applications demanding precise wavefront manipulation, a hybrid approach that combines RSS with the AoA technique can be adopted to enhance localization performance. In more detail, this approach discerns LoS components by their unique intensity and reception angles through the receiver's photodetectors (PDs) to detect variations in signal strength and arrival times due to the unique paths that signals follow from an LED to each detector. Consequently, by calculating the AoA based on the spatial distribution of the PDs, precise UE localization is enabled by mitigating the effects of non-LoS signals even in scenarios where only a single LED maintains LoS with the UE. However, it should be highlighted that this localization method requires equipping the receiver with multiple PDs to create angular diversity in the receiver's PD array.

To further elucidate the capabilities of the proposed PWE architecture, we refer to Fig.\ref{PWE_figure} and delineate the discrete scenarios involving the depicted UEs, showcasing the adaptability of the proposed architecture to diverse localization needs within the PWE. In more detail, localization of UE 1 is attained via the ceiling LEDs, while the position of UE 2  is determined through the LERIS, illustrating the system's innovative approach to utilizing LERIS for areas not covered by ceiling LEDs. Additionally, in scenarios where a UE is illuminated by more than 4 LEDs, such as UE 3 and UE 4, the architecture's flexibility is demonstrated as the UE selects the four highest RSS values to achieve precise localization, leveraging the strength of multiple optical anchors. Finally, UE 5, experiencing a LoS channel with only a single LED, requires the application of the hybrid RSS/AoA method, emphasizing the architecture's capability to adaptively employ different localization techniques based on the UE's specific situation. This versatility highlights the ability of the proposed PWE in Fig. 1 to accurately localize users under different conditions, validating its effectiveness in adapting to the dynamic and complex requirements of 6G networks.

\subsection{Optical Anchor-based LATC Scheme}

After integrating optical anchors into PWEs, it becomes feasible to offer enhanced PWE services through an optical anchor-based LATC scheme. This innovative approach takes advantage of the unique characteristics of optical anchors to ensure precise localization and seamless interaction between the UE and the PWE, unlocking the full potential of PWEs in the evolving context of 6G networks. This strategy not only ensures optimal service delivery but also provides a solid foundation for the cooperative operation of RIS components and optical anchors, thereby increasing the overall efficiency and responsiveness of the PWE system.

The proposed LATC scheme presented in Algorithm 1 starts with the AP broadcasting a beacon signal to announce the availability of the PWE, prompting UEs within it to select their preferred services and define the necessary QoS parameters. This initiates the localization process, where UEs calculate the number $N$ of LoS optical anchors available to them by analyzing received optical signals, which involves interpreting the unique intensity profiles of each optical anchor, a critical step in selecting the most appropriate localization method. When $N < 4$, meaning that there are not enough LoS paths for complete RSS localization, the localization error can become large. In this case, the UE employs the combined RSS/AoA method if the receiver is equipped with more than one PD, ensuring precise localization \cite{8113545}. In contrast, if $N \geq 4$, the UE is presented with the option to utilize either plain RSS or again a combination of RSS and AoA to achieve the desired level of localization precision as dictated by its chosen QoS. Subsequently, the UE reports its location $p$ to the AP, using the available RISs configured in diffusion mode to ensure the broad and uniform transmission of signals across the PWE, without requiring prior knowledge of the UE location. Despite the potential absence of a direct LoS channel between the UE and the AP, resulting in a weak RF transmission, the quantity of transmitted data is minimal, consequently, it can be transferred without errors with high probability. Upon receiving this information, the PWE meticulously selects the optimal RIS configuration through the available codebooks to match the reported location and the requested services.

\begin{algorithm}[t]

\linespread{0.85}\selectfont
\begin{algorithmic}[1]\label{alg1}
\caption{LATC Scheme based on Optical-Anchors} 
\State {Broadcast message to UEs advertising the existence of PWE}
\State{Select the desired PWE service and the required QoS}
\State {Calculate number of optical anchors $N$ that can serve each UE}
\For{Each UE}
\If{$N < 4$}
    \State {Employ the RSS/AoA method}.
\Else
    \State{Choose the 4 higher RSS values.}
    \State {Choose localization method based on QoS}.
\EndIf
\EndFor
\State Set RISs for diffusion.
\State Transmit UE location $p$ to AP.
\State Dynamically select optimal RIS configuration using $p$ and requested service.
\end{algorithmic}
\end{algorithm}

Following the initial configuration, the PWE can employ a dynamic feedback loop designed for real-time optimization, continuously adapting to changes in the UE position or the environment. This mechanism involves regular recalibration of the optical anchor setup and periodic updates to the codebook based on collected performance metrics. Such adjustments ensure the PWE remains responsive to dynamic user requirements and environmental variations, maintaining high levels of service quality and network performance. 

\subsection{Preliminary Results} 

As we explore enhanced electromagnetic functionalities, the importance of increasing the number of reflecting elements, $M$, becomes evident to improve the accuracy of steering electromagnetic waves towards the UE, crucial for applications requiring increased reliability. In this direction, Fig. 2 illustrates the variation in the scattering diagram of an RIS performing beam-steering for different numbers of reflecting elements. As can be observed, the directivity of the RIS increases with $M$, so that the half power beam width (HPBW) $\theta$ decreases from $18^o$ for an RIS with 50 elements to $10^o$ for 100 elements, and further to $2^{o}$ for an RIS with 1600 elements. This observation underscores the need for high-precision LATC schemes within PWEs since as the required RIS capabilities become more detailed, the margin for localization errors decreases significantly. To further emphasize this point, Fig. 3 illustrates the tolerated localization error $\sigma_p$ that we can have for the UE to be within $\theta$, where $\sigma_p$ equals the product of $\mathrm{tan}\left(\frac{\theta}{2}\right)$ and the RIS-UE distance. As shown, $\sigma_p$ diminishes with more reflecting elements, emphasizing the critical role of precise localization in PWEs to capitalize on the advanced capabilities of larger RISs, ensuring accurate delivery of RIS electromagnetic functions to the intended UEs.

Considering the pursuit of ultra-precise localization within PWEs, Fig. 4 shows the relationship between the localization error of the optical anchor-based RSS method and the parameter $K$, which represents the ratio of the power of the LoS component to the non-LoS components within the optical channels, over different Lambertian emission orders, $m$ of the optical anchors. Notably, given that $K$ typically exceeds 100 in practical optical channels, this method demonstrates an outstandingly low localization error, exceeding the precision required even for RIS configurations with up to 1600 reflecting elements. This result strongly suggests that optical anchor-based localization can meet the stringent localization accuracy requirements of sophisticated RIS functionalities and highlights its role in shaping PWEs. Moreover, it can be seen that increasing the directivity of optical anchors further refines the localization accuracy, however, for high values of $m$, we anticipate a decline in performance because users are likely to be outside this ultra-directive beam and experience low RSS values. Furthermore, this improvement comes at the cost of reducing the illuminated areas, highlighting the need to balance anchor directivity and localization performance to optimize the PWE performance.

  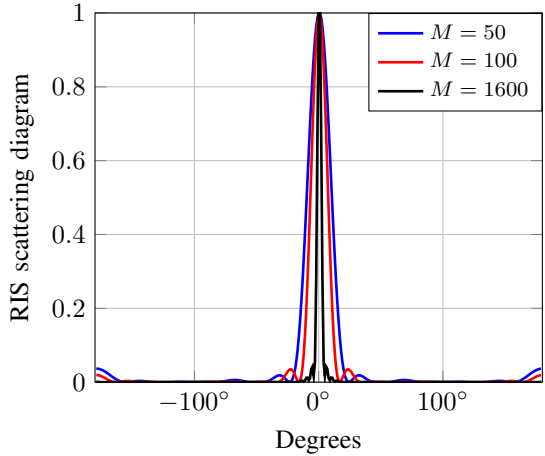
\begin{figure}
    \centering
    \begin{tikzpicture}
           \begin{axis}[
            width=0.85\linewidth,
            xlabel = {Degrees},
            ylabel = {RIS scattering diagram},
            ymin = 0,
            ymax = 1,
            xmin = -180,
            xmax = 180,
            xticklabel={$\pgfmathprintnumber{\tick}^\circ$},
            grid = major,
		legend entries ={{$M=50$},{$M=100$},{$M=1600$}},
            legend cell align = {left},
            legend style={font=\footnotesize},
            legend style={at={(1,1)},anchor=north east},
            ]
            \addplot[
            blue,
            no marks,
            line width = 1pt,
            style = solid,
            ]
            table {figs/fig1_1.dat};
            \addplot[
            red,
            no marks,
            line width = 1pt,
            style = solid,
            ]
            table {figs/fig1_100.dat};
            \addplot[
            black,
            no marks,
            line width = 1pt,
            style = solid,
            ]
            table {figs/fig1_1600.dat};
        \end{axis}
    \end{tikzpicture}
    \caption{Beam steering directivity versus number of reflecting elements}
    \label{fig:figure1}
  \end{figure}
  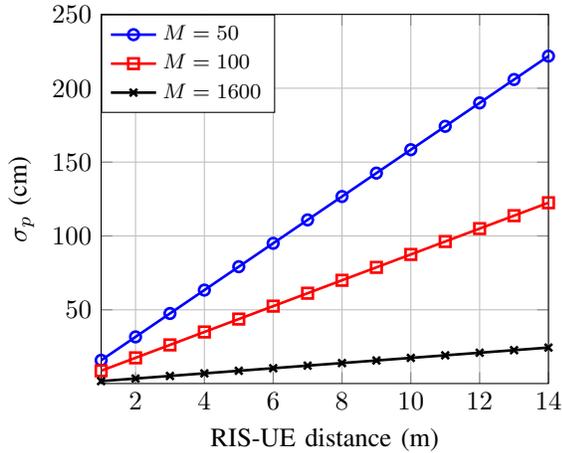
\begin{figure}
    \centering
    \begin{tikzpicture}
        \begin{axis}[
            width=0.85\linewidth,
            xlabel = {RIS-UE distance (m)},
            ylabel = {$\sigma_p$ (cm)},
            ymin = 0.1,
            ymax = 250,
            xmin = 1,
            xmax = 14,
            ytick = {0,50,...,250},
            grid = major,
            y filter/.code={\pgfmathparse{\pgfmathresult*100}}, 
		legend entries ={{$M=50$},{$M=100$},{$M=1600$}},
            legend cell align = {left},
            legend style={font=\footnotesize},
            legend style={at={(0,1)},anchor=north west},
            ]
            \addplot[
            blue,
            mark = o,
            mark repeat = 1,
            mark size = 2,
            line width = 1pt,
            style = solid,
            ]
            table {figs/fig2_50.dat};
            \addplot[
            red,
            mark = square,
            mark repeat = 1,
            mark size = 2,
            line width = 1pt,
            style = solid,
            ]
            table {figs/fig2_100.dat};
             \addplot[
            black,
            mark = x,
            mark repeat = 1,
            mark size = 2,
            line width = 1pt,
            style = solid,
            ]
            table {figs/fig2_1600.dat};
        \end{axis}
    \end{tikzpicture}
    \caption{Tolerated localization error $\sigma_p$ versus RIS-UE distance}
    \label{fig:figure2}
  \end{figure}
  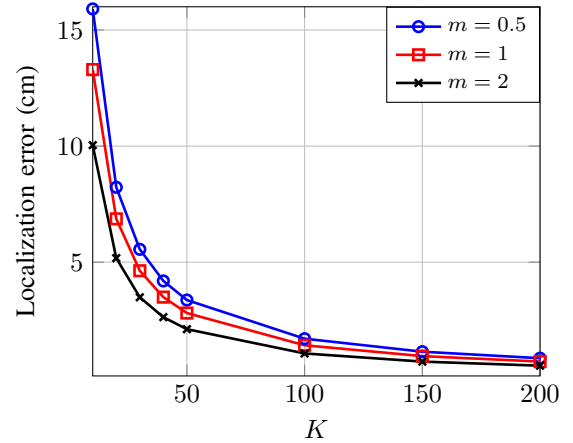
\begin{figure}
    \centering
    \begin{tikzpicture}
        \begin{axis}[
            width=0.85\linewidth,
            xlabel = {$K$},
            ylabel = {Localization error (cm)},
            ymin = 0.1,
            ymax = 16,
            xmin = 10,
            xmax = 200,
            grid = major,
		legend entries ={{$m=0.5$},{$m=1$},{$m=2$}},
            legend cell align = {left},
            legend style={font=\footnotesize},
            legend style={at={(1,1)},anchor=north east},
            ]
            \addplot[
            blue,
            mark = o,
            mark repeat = 1,
            mark size = 2,
            line width = 1pt,
            style = solid,
            ]
            table {m05.dat};
            \addplot[
            red,
            mark = square,
            mark repeat = 1,
            mark size = 2,
            line width = 1pt,
            style = solid,
            ]
            table {m1.dat};
             \addplot[
            black,
            mark = x,
            mark repeat = 1,
            mark size = 2,
            line width = 1pt,
            style = solid,
            ]
            table {m2.dat};
        \end{axis}
    \end{tikzpicture}
    \caption{Localization error of optical anchor-based RSS method versus $K$}
    \label{fig:figure3}
\end{figure}

\section{Challenges and Future Directions}\label{sec:challenges}

In the context of PWEs, the introduction of optical anchors presents both promising opportunities and significant challenges. This section explores the many challenges of seamlessly integrating optical anchors and provides insights into current limitations and potential avenues for future advancements. 

\vspace{-0.3cm}
\subsection{Challenges}
In the pursuit of efficiency within LED-based localization systems for PWEs, the placement of LEDs is crucial to ensure the pervasive presence of LoS signals throughout the PWE. This task entails a challenging optimization, considering multiple parameters, such as the field of view of the UEs. The situation is complicated further in scenarios lacking orientation extraction capabilities at the UE, demanding an LED deployment that provides strong LoS channels and avoids symmetries which could potentially complicate the location extraction, thereby offering both coverage and orientation information. Additionally, the integration of LEDs into the LERIS framework requires careful consideration of their compatibility with the metasurface, aiming to harmonize optical emissions with RF wave manipulation. This includes implementing advanced power routing protocols and thermal management strategies that are essential to dissipate the heat generated by the LEDs, ensuring that the heat emission does not compromise operational efficiency. By addressing these issues to fine-tune LED positioning for optimal LoS coverage, inaccuracies in UE localization that might result in suboptimal RIS configurations are minimized, ensuring the PWE maintains its service quality.

Furthermore, misalignments in the placement of RIS elements can profoundly impact performance, particularly in beamforming applications requiring ultra-directivity. Since codebooks are meticulously crafted for specific RIS element locations, any deviation from these prescribed positions introduces a significant mismatch between the anticipated and actual signal paths. This misalignment leads to substantial losses in beamforming efficacy, as the beams fail to converge optimally toward the intended targets. To address such disparities, rigorous calibration procedures become imperative. Through calibration, the system can adapt to the actual deployment conditions, compensating for misalignments and optimizing beamforming performance. Thus, meticulous attention to RIS element placement alongside comprehensive calibration processes is paramount to harnessing the full potential of beamforming technologies, ensuring optimal performance within the PWEs.

\vspace{-0.3cm}
\subsection{Future Directions}
Despite current challenges, future directions aim to enhance system performance in line with 6G network demands, particularly through the synergy of RF and optical technologies, which opens avenues for improved localization precision in PWEs. Specifically, RF signals offer wide coverage and robustness to obstructions, while optical signals, with their finer spatial resolution and immunity to RF interference, can be utilized for both communication and localization, thus facilitating cost-effective solutions for future networks \cite{9941537} and laying the groundwork for hybrid localization systems. Finally, integrating these signals via sensor fusion techniques alongside developing new RIS configurations tailored to the LED presence in the PWE, boosts the localization process centered on optical anchors. This comprehensive strategy ensures smooth transitions between localization schemes, fitting specific environmental needs or application demands, and highlights the combined potential of RF and optical frequencies in enhancing PWE functionalities.

Furthermore, central to the evolution of PWEs is the capabilities of AI to transcend current localization limitations. Specifically, AI can integrate RSS values from optical anchors with spatial data into a continuously updated spatial database, thus aligning RSS metrics with exact coordinates. This approach is notably effective given the LoS nature of optical frequencies, ensuring stable RSS values across locations and thus boosting PWE efficiency with optical anchors. Additionally, by utilizing advanced AI algorithms and the aforementioned spatial database, we can extract spatial information from non-illuminated areas, referred to as zero-power zones, and potentially foresee object movements or presences by extending mapping even to non-illuminated areas. Finally, exploiting historical data information in the proposed localization methods could significantly enhance performance, improving accuracy and reducing system complexity by narrowing the search domain, which reduces computational needs and maximizes efficiency and resource usage.

Finally, within PWEs, the imperative to accurately determine device positions without requiring active participation from UE necessitates the development of passive localization techniques. This need is particularly critical to support precise localization for devices with limited computational capabilities, such as IoT nodes. The complexity of this task is magnified by the challenge of discretizing the UE's antenna, further compounded by the demand for smaller antennas to align with the evolving requirements of 6G networks. In response to this challenge, imaging techniques through waveform manipulation emerge as a viable solution. By harnessing advanced signal processing techniques like waveform shaping and beamforming, these methods offer a pathway to achieving precise localization goals. Crucially, these techniques enable passive and simultaneous localization, ensuring seamless integration and optimal performance of PWEs across diverse UE scenarios.

\vspace{-0.5cm}
\section{Conclusion} \label{sec:conclusion}
In this paper, we have presented an in-depth investigation into the synergy between RIS and optical anchors within PWEs, aimed at achieving the high precision in localization required for unlocking the full potential of 6G network functionalities. Central to our exploration is the introduction of a newly proposed PWE architecture, equipped with multiple LEDs positioned on the ceiling and multiple LERISs, a design aimed at maximizing LoS signal coverage and enhancing localization accuracy across the PWE. Our findings underline the significance of precise localization, brought to the forefront by the integration of optical anchors, in enhancing RIS effectiveness and, by extension, the overall performance of PWEs. This exploration not only contributes to the theoretical advancement of wireless networks but also provides a practical framework for future research in the field, paving the way for the realization of 6G network standards and beyond.

\bibliographystyle{IEEEtran}
\bibliography{refs}

\end{document}